\documentclass[journal]{IEEEtran}
\IEEEoverridecommandlockouts

\usepackage{cite}
\usepackage{amsmath,amssymb,amsfonts}
\usepackage{booktabs}
\usepackage{algorithmic}
\usepackage{graphicx}
\usepackage{textcomp}
\usepackage{xcolor}
\usepackage{latexsym}
\usepackage [english]{babel}
\usepackage [autostyle, english = american]{csquotes}
\usepackage{placeins}

\MakeOuterQuote{"}
\def\BibTeX{{\rm B\kern-.05em{\sc i\kern-.025em b}\kern-.08em
    T\kern-.1667em\lower.7ex\hbox{E}\kern-.125emX}}
\begin{document}

\title{Diffusion Dynamics in Biofilms with \\Time-Varying
Channels}

\author{\IEEEauthorblockN{Yanahan Paramalingam and Adam Noel, \IEEEmembership{Senior Member,~IEEE}}

\thanks{For the purpose of open access, the authors have applied a Creative Commons Attribution (CC-BY) licence to any Author Accepted Manuscript version arising from this submission.} 

\thanks{Y.~Paramalingam is with the School of Engineering, University of Warwick, Coventry, UK. (e-mail: \{yanahan.paramalingam.1@warwick.ac.uk)}

\thanks{A.~Noel is with the Department of Electrical and Computer Engineering
at Memorial University, St. John's, Canada. (e-mail: adam.noel@mun.ca)}
}

\maketitle

\begin{abstract}

A biofilm is a self-contained community of bacteria that uses signaling molecules called autoinducers (AIs) to coordinate responses through the process of quorum sensing.  Biofilms exhibit a dual role that drives interest in both combating antimicrobial resistance (AMR) and leveraging their potential in bioprocessing, since their products can have commercial potential. Previous work has demonstrated how the distinct anisotropic channel geometry in some biofilms affects AIs propagation therein. In this paper, a 2D anisotropic biofilm channel model is extended to be a time-varying channel (TVC), in order to represent the diffusion dynamics during the maturation phase when water channels develop. Since maturation is associated with the development of anisotropy, the time-varying model captures the shift from isotropic to anisotropic diffusion. Particle-based simulation results illustrate how the TVC is a hybrid scenario incorporating propagation features of both isotropic and anisotropic diffusion. This hybrid behavior aligns with biofilm maturation. Further study of the TVC includes characterization of the mutual information (MI), which  reveals that an increased AI count, reduced transmitter -- receiver distance, greater degree of anisotropy, and shorter inter-symbol interference lengths increase the MI. Finally, a brief dimensional analysis demonstrates the scalability of the  anisotropic channel results for larger biofilms and timescales.

\end{abstract}


\section{Introduction}

Biofilms are microbial cities that consist of bacteria, extracellular DNA (eDNA), proteins, and polysaccharides embedded in an extracellular polymeric substance (EPS) \cite{hall2004bacterial, costerton1999bacterial, vestby2020bacterial}.  The communication process of quorum sensing (QS) is facilitated by signalling molecules known as autoinducers (AIs). AIs are vital for the survival of a biofilm, as they enable bacteria to regulate key processes such as population density and species composition \cite{mukherjee2019bacterial}. In biofilms, water channels are utilised as an efficient transportation system for AIs, nutrients, and waste \cite{quan2022water,wilking2013liquid,rooney2020intra}. With an increase in efforts to produce valuable biofilm byproducts \cite{rosero2021microbial} and combat AMR \cite{bjarnsholt2007quorum,paluch2020prevention}, there has been increasing interest in the scientific community to produce robust mathematical models of biofilm signaling and behavioral dynamics \cite{kannan2018mathematical,perez2016mathematical,li2020quantitative,mattei2018continuum}.  Furthermore, the development of a sufficiently robust model includes key challenges that need to be addressed, including  dynamic environments, complex QS mechanisms, and external stress factors. Advancements in both modeling and experimental methods can help  to address these challenges \cite{paramalingam2025anisotropic}.

The field of molecular communication (MC) has excellent potential to advance communication model development for biofilms due to its suitability for characterizing AI propagation \cite{farsad2016comprehensive}. The standard diffusion model employed in MC is isotropic (i.e., uniform) diffusion. However, to model biofilms, anisotropic (i.e., non-uniform) diffusion should be considered, particularly in biofilms that have water channels \cite{van2012anisotropic}. Bacterial communication has been widely studied in MC \cite{noel2017effect,martins2022microfluidic,einolghozati2013relaying,balasubramaniam2023realizing,cobo2010bacteria,unluturk2015genetically,tissera2020bio,einolghozati2013design,fang2020characterization,abadal2012quorum,tissera2019quorum,li2015evaluation}, including consideration of QS disruption in biofilms \cite{martins2018molecular,michelusi2016queuing,martins2016using,gulec2023stochastic}. In \cite{paramalingam2025anisotropic}, we were the first to introduce an anisotropic diffusion model for molecular communication in biofilm and further investigated different placements of point-to-point transmitter (TX) and receiver (RX) links in bounded 2D space. Under our proposed model, the diffusion coefficient along the axial direction (i.e., along the curvature of the biofilm) was lower than the diffusion coefficient along the radial direction (i.e., running through the center of the biofilm). We derived the channel impulse response (CIR) using Green’s function for concentration (GFC) and validated it with particle-based simulation (PBS). Our findings demonstrated that when the TX was positioned away from the center of the  biofilm, there were higher diffusion peaks under the anisotropic model, with the propagation travel time of AIs being inversely proportional to both overall biofilm size and diffusion coefficient values. This supported the hypothesis that signals propagate faster from the biofilm edge to the center when the channel is anisotropic, thereby helping bacteria in a mature biofilm to respond faster to environmental changes.

Our previous anisotropic diffusion model assumed a mature biofilm under ideal conditions including constant temperature, constant pH, and the exclusion of other external environmental influences, and without considering the impact of biofilm growth on the diffusion dynamics over time, i.e.,  the system geometry was fixed. However, during the maturation phase, the formation of water channels could be modeled as a propagation channel that becomes gradually more anisotropic over time, such that the corresponding diffusion coefficients are also time-varying. Biofilm maturation is driven by local chemical gradients that increase EPS secretion, leading to an intricate network of rigid water channels. The water channels provide an enhanced pathway for nutrient and oxygen transport, as well as the removal of metabolic waste. The improved architecture makes a biofilm better able to withstand environmental stresses \cite{flemming2010biofilm,kuchma2000surface}.

To account for the impact of water channel formation on signal propagation during maturation, we propose considering a time-varying channel (TVC) model where the diffusion coefficients become functions of time. Furthermore, since water channel formation is a gradual process (i.e., on the scale of hours \cite{wilking2013liquid}), we propose that it is sufficient to model the TVC as a channel whose diffusion coefficients change between consecutive time intervals. The inclusion of TVC in the previously-introduced anisotropic propagation model and demonstrating its impact on the propagation of AIs as the biofilm matures can improve the model's suitability for describing and predicting the role of QS.

Other prior investigations have incorporated TVC into MC models to produce more robust biological models of communication. In targeted drug-delivery, mobile nanomachines (acting as the  TX or RX) have been modeled using Brownian motion to obtain a time-dependent CIR \cite{ahmadzadeh2018stochastic, 8254237,cao2019diffusive,rouzegar2019diffusive}. Similarly, to mimic limited  resource availability in energy-constrained models  (e.g., due to a cell's finite metabolic usage), molecule usage under TVC conditions was optimized in \cite{feng2018resource}. TVC has also been considered in anisotropic \cite{trinh2022molecular} and advection \cite{lin2012signal} models that could have applications in blood vessels or tissue layers. While these previous works are related to our modeling due to their consideration of TVC conditions, to the best of our knowledge, no previous MC work has considered TVC in a biofilm.

The aim of this paper is to build on the bounded 2D system model for anisotropic diffusion in a biofilm,  as presented in \cite{paramalingam2025anisotropic}, to incorporate a TVC to better capture the evaluation of diffusion dynamics as the biofilm water channels mature. The main contributions of this paper are summarized as follows:

\begin{enumerate}
    \item We introduce a time-varying channel (TVC) to the anisotropic diffusion model that we developed in \cite{paramalingam2025anisotropic} to represent the evolving AI diffusion dynamics during the biofilm's maturation phase when the water channels develop. The TVC is designed as a series of time intervals with fixed geometry, each represented by a coherence time with fixed radial and axial diffusion coefficients. In principal, the TVC could take on any diffusion coefficient values within a given interval, but we focus on the practical case where the biofilm is initially isotropic and gradually becomes more anisotropic as its water channels mature.
    \item We  use particle-based simulations (PBS) to generate colormaps that compare the diffusion dynamics across isotropic, anisotropic, and TVC models. Observations of the local peak concentrations and the times at which they occur reveal that the anisotropic model consistently produces the earliest and highest peaks, reflecting that this model has a more directed, quasi-1D diffusion behavior. In contrast, the TVC exhibits hybrid characteristics that capture features of both isotropic and anisotropic diffusion as the system evolves over time.
    \item We perform a case study with dimensional analysis principles to demonstrate the scalability of our model. The case study shows how our results can be applied to larger biofilms and over longer timescales than those that we simulate directly, even when the diffusion is anisotropic.
    \item We characterize the potential communications performance in the TVC by measuring the mutual information  (MI) between a sequence of ON-OFF keying symbols encoded at a TX and  the corresponding  noisy observations at a RX. We study the sensitivity of the MI to varying system parameters,  such as the number of AIs released per symbol and the TX-RX distance.

\end{enumerate}

The rest of the paper is organized as follows. Section II summarizes the analytical model for anisotropic diffusion in a biofilm and presents the performance criteria that are used to assess  the different models. In Section III, we present and discuss all of the simulation results. Finally, Section IV  concludes the paper.

\section{Analytical Modeling}

This section presents the analytical modeling that we apply in this paper to characterize point-to-point transmission via time-varying anisotropic diffusion between a TX and RX placed arbitrarily within a bounded 2D biofilm. First, we review the corresponding expected channel impulse response  for static anisotropic diffusion, based on the closed-form expression of the Green's function for concentration, which we previously derived in \cite{paramalingam2025anisotropic}. Then, we take the derivative of the CIR with respect to time to derive the expected time of the peak concentration observed at the RX for the static case. To extend the physical model to the time-varying case, we vary the axial diffusion coefficient over a sequence of intervals. Finally, we describe the transmission scheme between the TX and RX that we use to calculate the mutual information.

\subsection{Static Anisotropic Diffusion Model}

\begin{figure}[t!]
\includegraphics[width=1\linewidth]{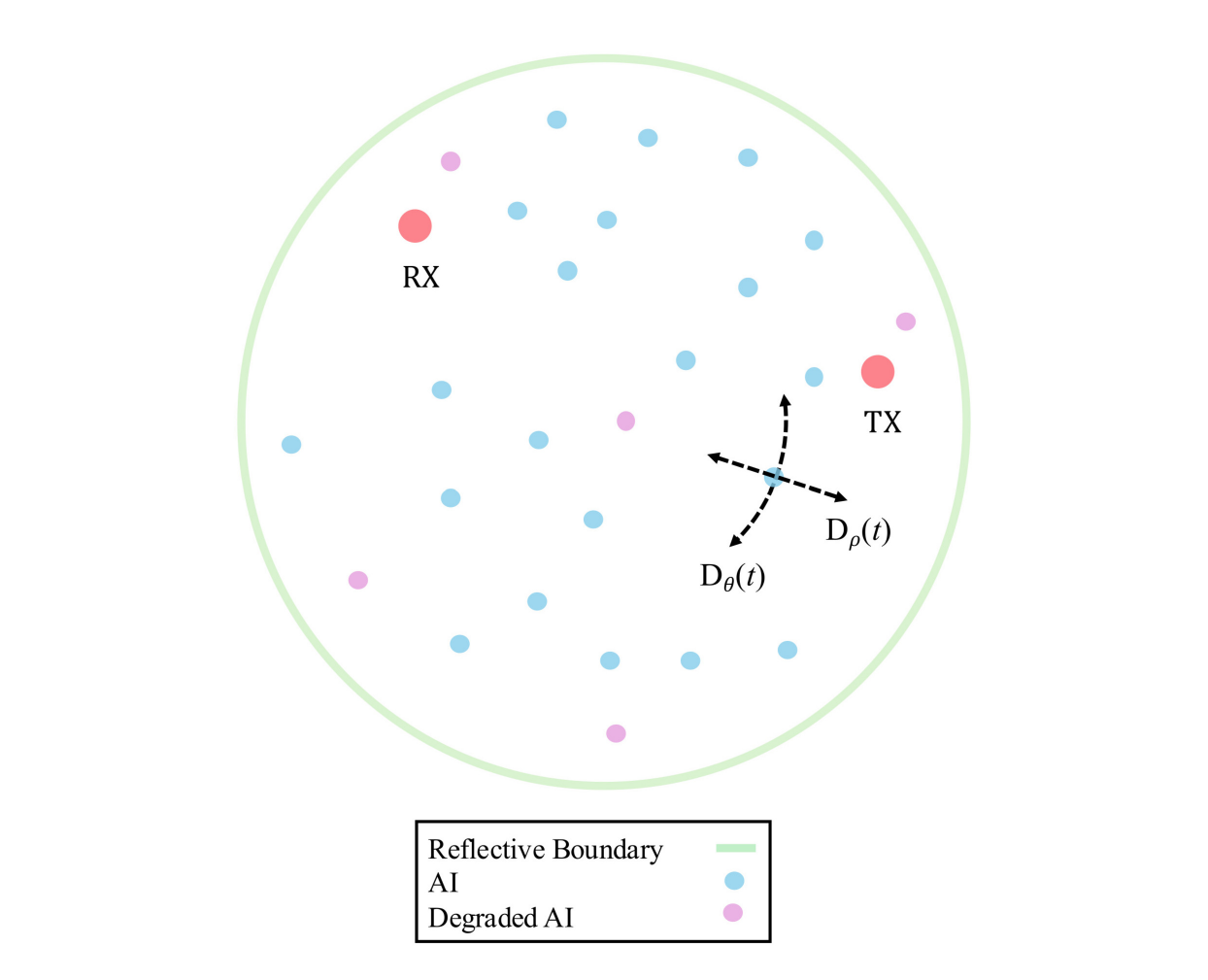}
\caption{Schematic representation of AI molecule propagation in a 2D biofilm with a time-varying channel (TVC). There is anisotropic propagation from a point TX to a passive and transparent RX. AIs diffuse according to $D_{\rho}(t)$ and $D_{\theta}(t)$, which are both constant in the static channel case. The outer circular boundary is reflective and the AIs can degrade according a first-order degradation process.}
\label{System Model Diagram}
\end{figure}

The anisotropic diffusion model that we developed in \cite{paramalingam2025anisotropic} assumed that molecules propagated within a mature biofilm, as informed by [9]–[11]. For such a biofilm, we can assume that its size is constant and the layout of its water channels are fixed. AI molecules are able to diffuse along the water channels, which are primarily aligned along the radial direction towards the center of the  biofilm, but they can also diffuse around bacteria through the EPS. Thus, AIs diffusing through the biofilm but not along a water channel are in effect propagating through a porous medium. Instead of separately  modeling propagation through the water channels and EPS, in \cite{paramalingam2025anisotropic} we modeled the biofilm as a homogeneous anisotropic diffusion channel with constant pH  and temperature. By defining a polar coordinate system with the origin at the center of the biofilm and with (\(\rho\), \(\theta\)) denoting radial and azimuthal coordinates, respectively, we characterized the diffusion with distinct radial and azimuthal diffusion coefficients $D_{\rho}$ and $D_{\theta}$, respectively (see Fig.~\ref{System Model Diagram}). Since water channels should facilitate faster propagation than the denser porous regions, we approximated biofilm diffusion by setting radial diffusion to be faster than axial diffusion (i.e., $D_{\rho} > D_{\theta}$).

The simplified system is within a circle of radius $\rho_{c}$. The circle has a fully-reflecting outer boundary, no molecule sources besides the TX, no local molecule sinks, and no external environmental influences. The TX is an impulsive point source at $\Bar{r}\textsubscript{tx}$ = ($\rho\textsubscript{tx}$, $\theta\textsubscript{tx}$),
where 0 $\leq \rho\textsubscript{tx} \leq \rho_{c}$. The RX is a passive and transparent observer that is able to count the  number of AIs therein. Once AIs are released by the TX, they are able to degrade, be consumed, or transform into another molecule at a certain lumped rate, \(k_d\) $s^{-1}$. We approximate this conversion as a first-order degradation reaction, i.e.,
\begin{equation}
    A \xrightarrow{k_d}\phi.
    \label{equation 1}
\end{equation}

We characterize the impulsive TX molecule release at time $t\textsubscript{0}$ with the function $S(\Bar{r},t,\Bar{r}\textsubscript{tx},t\textsubscript{0}) = \delta\frac{(\rho-\rho\textsubscript{tx})}{\rho}\delta(\theta-\theta\textsubscript{tx})\delta(t-t\textsubscript{0})$. The concentration $C(\bar r,t|{\bar r_{\mathrm{ tx}}},{t\textsubscript{0}})$ is  then the  corresponding molecule concentration at point $\Bar{r}$ and  time ${t}>t\textsubscript{0}$. The anisotropic AI diffusion within the biofilm is then governed by the following partial differential equation (with the boundary condition $\frac{\partial C}{\partial \rho} = 0$ at the outer boundary)
\begin{multline}\label{equation 3}
\frac {{\partial C(\bar r,t|{\bar r_{\mathrm{ tx}}},{t\textsubscript{0}})}}{\partial t} = \mathbf{\nabla} \cdot \biggl(D\textsubscript{eff}\cdot \Vec{\mathbf \nabla}\ C(\bar r,t|{\bar r_{\mathrm{ tx}}},{t\textsubscript{0}})
\\
- k\textsubscript{d}C(\bar r,t|{\bar r_{\mathrm{ tx}}},{t\textsubscript{0}}) + S(\bar r,t|{\bar r_{\mathrm{ tx}}},{t\textsubscript{0}})\biggr),
\end{multline}
where
\begin{equation}
     D\textsubscript{eff} = \begin{bmatrix} D_{\rho}  & 0 \\0 &  D_{\theta}  \end{bmatrix}.
     \label{equation 4}
\end{equation}

In \cite{paramalingam2025anisotropic}, we proved that the solution to \eqref{equation 3} is
\begin{align}
C(\bar r,t) = & \sum_{n=0}^\infty\sum_{m=1}^\infty \biggl({\frac{{L_{n}{J_{\zeta}}\left(\lambda_{nm}\rho\textsubscript{tx}\right)}}{{{N_{nm}}}}} \nonumber \\
& \times{J_{\zeta}}\left(\lambda_{nm}\rho\right)\cos(n(\theta-\theta_{tx})) \nonumber \\
& \times e^{-D_\rho \sqrt{\lambda_{nm}}(t - t\textsubscript{0})}u(t - t\textsubscript{0})\biggr),
\label{equation 5}
\end{align}
which also serves as the expected CIR between the TX and a receiver at $\Bar{r}$. In other words, \( C(\bar{r}, t) \) describes the time-dependent molecule concentration at the point \( \bar{r} = (\rho, \theta) \)  within the biofilm due to an instantaneous TX transmission at \( \bar{r}_{\text{tx}} = (\rho_{\text{tx}}, \theta_{\text{tx}}) \) at time $t\textsubscript{0}$.

Eq.~(4) represents the channel impulse response for anisotropic diffusion as a double summation over azimuthal and radial modes indexed by \(n < 3\) and \(m < 5\). The degree of anisotropy is captured by the Bessel function of the first kind, \(J_\zeta(\cdot)\), whose order is \(\zeta = \sqrt{D_\theta / D_\rho}\, n\). The eigenvalues \(\lambda_{nm}\) are the \(m\)th roots of the boundary condition \(D_\rho \lambda_{nm} J'_\zeta(\lambda_{nm}\rho_c) + k_f J_\zeta(\lambda_{nm}\rho_c) = 0\), which reduces to \(J'_\zeta(\lambda_{nm}\rho_c) = 0\) under perfectly reflective conditions (\(k_f = 0\)). The coefficients \(L_n\) arise from the angular delta-function expansion, with \(L_0 = 1/(2\pi)\) and \(L_n = 1/\pi\) for \(n \ge 1\). The normalization factor \(N_{nm} = \int_0^{\rho_c} \rho\, J_\zeta^2(\lambda_{nm}\rho)\, d\rho\) ensures orthogonality of the radial modes. The angular separation between the transmitter and receiver is described by \(\cos[n(\theta - \theta_{\mathrm{tx}})]\), while the temporal decay term \(\exp[-D_\rho\sqrt{\lambda_{nm}} (t - t_0)]\) captures the effects of diffusion and molecular degradation over time. Finally, Eq.~(\ref{equation 5}) also applies to the isotropic diffusion case by setting \(D_\rho = D_\theta\).

\subsection{Propagation Peak Time}
\label{sec_peak}
One of the metrics that we will use to compare the different channel models in this paper is the local peak AI concentration and time of peak concentration. Thus, we will find it insightful to evaluate the time-derivative of the CIR in (\ref{equation 5}). Fortunately, there is only one term in (\ref{equation 5}) that is a function of time \(t\) for \(t > t_0\). Using the property of the derivative of an exponential, we can immediately write the time derivative of the CIR for \(t > t_0\) as

\begin{align}
\frac{\partial C(\bar{r}, t)}{\partial t} & \sum_{n=0}^\infty\sum_{m=1}^\infty \biggl({\frac{{L_{n}{J_{\zeta}}\left(\lambda_{nm}\rho\textsubscript{tx}\right)}}{{{N_{nm}}}}} \nonumber \\
& \times{J_{\zeta}}\left(\lambda_{nm}\rho\right)\cos(n(\theta-\theta_{tx})\biggr) \nonumber \\
&\quad\times \Bigl[-D_{\rho}\sqrt{\lambda_{nm}}\,e^{-D_{\rho}\sqrt{\lambda_{nm}}(t - t\textsubscript{0})}\,u(t - t\textsubscript{0}) \nonumber\\
&\quad + e^{-D_{\rho}\sqrt{\lambda_{nm}}(t - t\textsubscript{0})}\,\delta(t - t\textsubscript{0})\Bigr]. \label{equation peak}
\end{align}

By setting the time-derivative (\ref{equation peak}) equal to zero, we can numerically evaluate the expected time at which the peak AI concentration is observed at some location $\bar{r}$. Then, we can substitute that time into (\ref{equation 5}) to calculate the expected concentration at that time and location.

\subsection{Time-Varying Anisotropic Diffusion Model}

Water channels within biofilms slowly develop and grow while the biofilm is also growing. Once the biofilm reaches its radial growth limit, e.g., due to limited space or nutrient availability, it is in a maturation phase where the water channels continue to grow and start to merge [9]–[11]. We now consider AI propagation during this phase, before the static anisotropic diffusion channel is established.  We characterize this period by enabling the diffusion  coefficients $D_{\rho}$ and $D_{\theta}$ to be time-varying.

We assume that the water channel merging process makes the overall biofilm diffusion environment more anisotropic. However, we also assume that this process is slow relative to the typical molecule propagation time \cite{wilking2013liquid}. Thus, we propose a simplified TVC model that assumes the existence of a channel coherence time $T$, i.e., an interval over which we can assume that the water channel geometry and bacterial EPS conditions remain fixed and the degree of anisotropy remains constant for the entire interval. Within a  given time interval  $T$, $D_{\rho}$ and $D_{\theta}$ remain constant. The TVC is characterized by a sequence of time intervals, each with their own corresponding diffusion coefficients. To simulate this channel,  we simply need to update the diffusion coefficients for each coherence time interval. In particular, we  assume that the biofilm gradually becomes more anisotropic with each subsequent interval as it matures.

\subsection{Mutual Information Computation with Inter-Symbol Interference}

The mutual information (MI) quantifies the dependence between the transmitted symbol 
\(X\) and the received signal \(Y\) (i.e., the AI count at the RX). The transmitted symbol is represented as a random variable \(X\), whose realization is  \(a_s\), in the \(s\)-th interval where \(s \ge 0\).

\subsubsection{Transmission Model}

The binary concentration shift keying (BCSK) modulation scheme is utilized to evaluate MI in the TVC. At the start of each symbol interval a pulse of AIs are emitted from the TX located at \(\bar{r}_{\mathrm{tx}}=(100~\mu\mathrm{m},\,0~\mathrm{rad})\), in order to convey the binary symbol \(a_s \in \{0,1\}\). If \(a_s=1\), then a fixed number of AIs \(N_1\) are released; if \(a_s=0\), then \(N_0\) AIs are released, with \(N_1 > N_0\). We utilize  On--Off Keying (OOK) where \(N_1=N\) and \(N_0=0\), thus, there is no emission for symbol ``0.'' The \textit{a priori} symbol probabilities are

\[
p_x = \Pr(X = x), \quad \text{with} \quad p_0 + p_1 = 1.
\]

\subsubsection{Symbol Intervals and Sampling}

In the system, time is divided into symbol intervals \([sT_s,(s+1)T_s)\) of duration \(T_s\). In each interval \(s\), an impulse emission occurs at \(t=sT_s\), and the RX takes a single measurement at \(t=(s+1)T_s\). With sampling step \(\Delta t\), let \(N_{\mathrm{sym}} \in \mathbb{N}\) be the number of samples per symbol so that \(T_s = N_{\mathrm{sym}}\Delta t\). Thus, emissions occur at samples \(k=sN_{\mathrm{sym}}\) and measurements at \(k=(s+1)N_{\mathrm{sym}}\).

The discrete-time emission is
\begin{equation}
M[k]=\sum_{s=0}^{L-1} N_s\,\delta\big[k-sN_{\mathrm{sym}}\big],
\label{eq:disc_emission}
\end{equation}
where \(\delta[\cdot]\) is the Kronecker delta, \(L\) denotes the total number of transmitted symbols in the sequence, and \(N_s\) is the number of AIs released during the the \(s\)-th symbol interval according to the transmitted symbol \(a_s\).  

\subsubsection{RX Observation and MI Estimation}

For each symbol interval \(s\), the AI count measured during that interval at RX is denoted \(Y\). At the RX, the observed AI count \(Y\) is measured once at the end of each symbol interval, at \(t = (s+1)T_s\) for symbol \(s\). To measure the mutual information, we first present the \textit{memoryless} case, i.e., for $W$ trials of \(s=0\) (even though we do not consider this case in our results). The mutual information between \(X\) and \(Y\) is then computed as
\begin{align}
I(X;Y)
&= \sum_{x \in \{0,1\}} p_x
   \sum_{y=0}^{y_{\max}} \hat{P}(Y=y \mid X=x) \nonumber\\
&\quad\times
   \log_2\!\frac{\hat{P}(Y=y \mid X=x)}{\hat{P}(Y=y)}.
\label{eq:MI}
\end{align}

The conditional probability mass function (PMF) \(\hat{P}(Y=y \mid X=x)\) represents the probability of observing AI count \(y\) given transmission of symbol \(x\). The empirical marginal PMF of \(Y\) is \(\hat{P}(Y=y)\), and \(y_{\max}\) denotes the maximum observed AI count across all $W$ trials, used as a practical summation limit.  

The conditional PMF is estimated as
\begin{equation}
\hat{P}(Y=y \mid X=x) = \frac{1}{|\mathcal{W}_x|} \sum_{w \in \mathcal{W}_x} \mathbb{I}(Y_w = y),
\label{eq:condPMF}
\end{equation}
\noindent where \(\mathcal{W}_x\) is the set of trials in which symbol \(x\) was transmitted, \(\mathbb{I}(\cdot)\) is the indicator function that equals 1 if the observed count in trial \(w\) was \(y\), $Y_w$ is the output in trial $w$, and \(|\mathcal{W}_x|\) is the number of such trials. The marginal PMF is then

\begin{equation*}
\hat{P}(Y=y) = p_0\,\hat{P}(Y=y \mid X=0)
\end{equation*}
\begin{equation}
+ p_1\,\hat{P}(Y=y \mid X=1).
\label{eq:margPMF}
\end{equation}


\subsubsection{Modeling MI Under Inter-Symbol Interference}

To consider intervals \(s>0\), we must consider memory and account for inter-symbol interference (ISI). For clarity, we denote the random variables for the transmitted symbol and received signal as \(X_s\) and \(Y_s\), respectively. ISI can take place since AIs from previously-transmitted symbols can persist in the biofilm and be present to be counted with current observations. Hence, the received signal for symbol index $s$ can be expressed as

\begin{equation}
Y_s = f(a_s, a_{s-1}, \ldots, a_{s-K}) + \eta_s,
\end{equation}
where \(K\) is the ISI length, \(f(\cdot)\) is the deterministic channel mapping that accounts for ISI, and the stochastic noise term capturing fluctuations is denoted by \(\eta_s\). When considering ISI, the received count in the \(s\)-th interval depends on both the past and current transmitted symbols. Consequently, the conditional distribution used in the MI computation becomes \(\hat{P}(Y_s \mid X_s, X_{s-1}, \ldots, X_{s-K})\). However, since we only compute mutual information directly from our simulation data, the ISI is inherently included for any $s>0$. In our results, we consider the transmission of 5-symbol sequences, so the largest $K$ that we can observe is $L-1=4$.

\subsubsection{Normalization}

To enable comparisons of MI, we normalize MI by the source entropy, i.e.,

\begin{equation}
I_{\text{norm}} = \frac{I(X;Y)}{H(X)},
\label{eq:MI_norm}
\end{equation}
where
\begin{equation}
H(X) = -p_0 \log_2 p_0 - p_1 \log_2 p_1.
\label{eq:entropy}
\end{equation}

This normalization ensures that \(I_{\text{norm}} \in [0,1]\), providing a dimensionless and interpretable measure of the effectiveness of AI propagation.

\section{Simulation and Numerical Results}

In this section, we present simulation results to assess the AI propagation in the time-varying channel (TVC). First, we demonstrate the dimensional scaling of the anisotropic diffusion channel with fixed geometry. This test is performed to justify simulating over relatively short timescales and distances with small molecule counts, since the results are scalable to larger systems. Next, we compare fixed-geometry channels (with either isotropic or anisotropic diffusion) with the anisotropic diffusion TVC. We consider both the spatiotemporal diffusion profile and the local diffusion peaks. Finally, we assess the mutual information within the TVC and how it is sensitive to different system parameters.

Since the channel impulse response of the static anisotropic channel, as presented in Eq.~(5), was previously validated in \cite{paramalingam2025anisotropic}, this section emphasizes results generated from a particle-based simulator (PBS) that we implemented in MATLAB (R2023a; the MathWorks, Natick, MA, USA).  The PBS simulates diffusion as independent random events by tracking the positions of the AI molecules in 2D using polar coordinates. The AIs are released from a point source and observed over a grid of locations that covers the entire biofilm. AI propagation is executed over discrete time intervals of $\Delta t$ and diffusion parameters are adjusted according to the channel coherence time $T$. The outer boundary is fully reflective for all AI molecules. We ignore AI degradation; i.e., we set $k_d=0\,\text{s}^{-1}$. Unless noted otherwise, we place the TX at the rightmost edge of the biofilm boundary (i.e., at $\rho=\rho_c$ and $\theta=0$), release AIs at time $t=0$, and simulate each set of system parameters once. The other default simulation parameters used are listed in Table~\ref{Table 1}.

\begin{table}[h!]
\centering 
\caption{Parameters for the proposed biophysical model} 
\label{tab:my_label} 
\resizebox{\columnwidth}{!}{
\begin{tabular}{@{}lr@{}} 
\toprule 
\textbf{Parameter} & \textbf{Value} \\
\midrule 
Diffusion coefficient in radial direction, \(D_{\rho}\) & \(5\times 10^{-10} \, \text{m}^2 \cdot \text{s}^{-1}\) \\
Diffusion coefficient in azimuth direction, \(D_{\theta}\) & \(\{5\times 10^{-10}, 5\times 10^{-11}\} \, \text{m}^2 \cdot \text{s}^{-1}\) \\
Circle radius, \(\rho _{c}\) & 100 \(\mu\text{m}\) \\
Point source transmitter location, \(\bar{r}_{\mathrm{tx}}\) & \(\{100\}\) \(\mu\text{m}\), 0 rad \\
Degradation reaction constant, \(k_d\) & 0 \(\text{s}^{-1}\) \\
Receiver radius & 1 \(\mu\text{m}\) \\
Number of transmitted molecules inside the biofilm $N$ & \(2 \times 10^{5}\) \\
Time step in PBS, \(\Delta t\) & \(10^{-1}\)\,s \\
Number of time steps in PBS & 800 \\
TVC coherence time, $T$ & $20\,\text{s}$\\
Symbol interval, \(T_s \)  & $20\,\text{s}$\\
Color map pixel element width, $\ell$ & $2.5\,\mu$m \\
\bottomrule 
\end{tabular}
}
\label{Table 1}
\end{table}

We focus on fixed radial diffusion \( D_{\rho} = 5\times 10^{-10} \, \text{m}^2 \cdot \text{s}^{-1} \), whose value was derived from experimental data of isotropic AHL molecule diffusion within biofilms \cite{alberghini2009consequences}. To demonstrate the TVC during the biofilm maturation phase, we varied \( D_{\theta} \) from $5 \times 10^{-10}$ to $5 \times 10^{-11}\,\text{m}^2 \cdot \text{s}^{-1}$. In all circumstances, $D_{\theta} \le D_{\rho}$ to reflect the reduced diffusion within the EPS relative to the water channels. More specifically, we make the TVC more anisotropic as the biofilm matures, as follows:

\begin{equation}
D_{\theta}(t) =
\begin{cases} 
5\times 10^{-10}\,\text{m}^2 \cdot \text{s}^{-1}, & 0 \le t \leq 20\,\text{s} \\
1 \times 10^{-10}\,\text{m}^2 \cdot \text{s}^{-1}, &  20\,\text{s} < t \leq 40\,\text{s} \\
7 \times 10^{-11}\,\text{m}^2 \cdot \text{s}^{-1}, &  40\,\text{s}< t \leq 60\,\text{s} \\
5 \times 10^{-11}\,\text{m}^2 \cdot \text{s}^{-1}, &  t > 60\,\text{s}.
\end{cases}
\label{eqn_timevarying}
\end{equation}

Many of our results are plotted using color maps. Every color map is drawn with a white circle indicating the outer boundary of the biofilm. Each pixel represents a square patch and the corresponding color reflects the observed value at that location, i.e., the number of AIs present (used as a proxy for concentration) or the time at which the peak number of molecules was observed. The default square width is $\ell=2.5\,\mu$m.

\subsection{Dimensional Scaling Under Anisotropic Diffusion}

Our first test is to confirm the dimensional scaling of diffusion in our anisotropic channel. Our default system parameters consider a biofilm with a radius of \(\rho _{c} = 100\,\mu\text{m}\) and most of our tests measure diffusion for only $80\,$s. The scalability of isotropic diffusion is well-known \cite{Noel2013b}; the propagation time for a diffusion wave released from a point-source into an unbounded medium increases with the square of the distance. Such intuition has not yet been established for our anisotropic channel which has different diffusion rates in the radial and axial directions.

\begin{figure}[t!]
\includegraphics[width=1\linewidth]{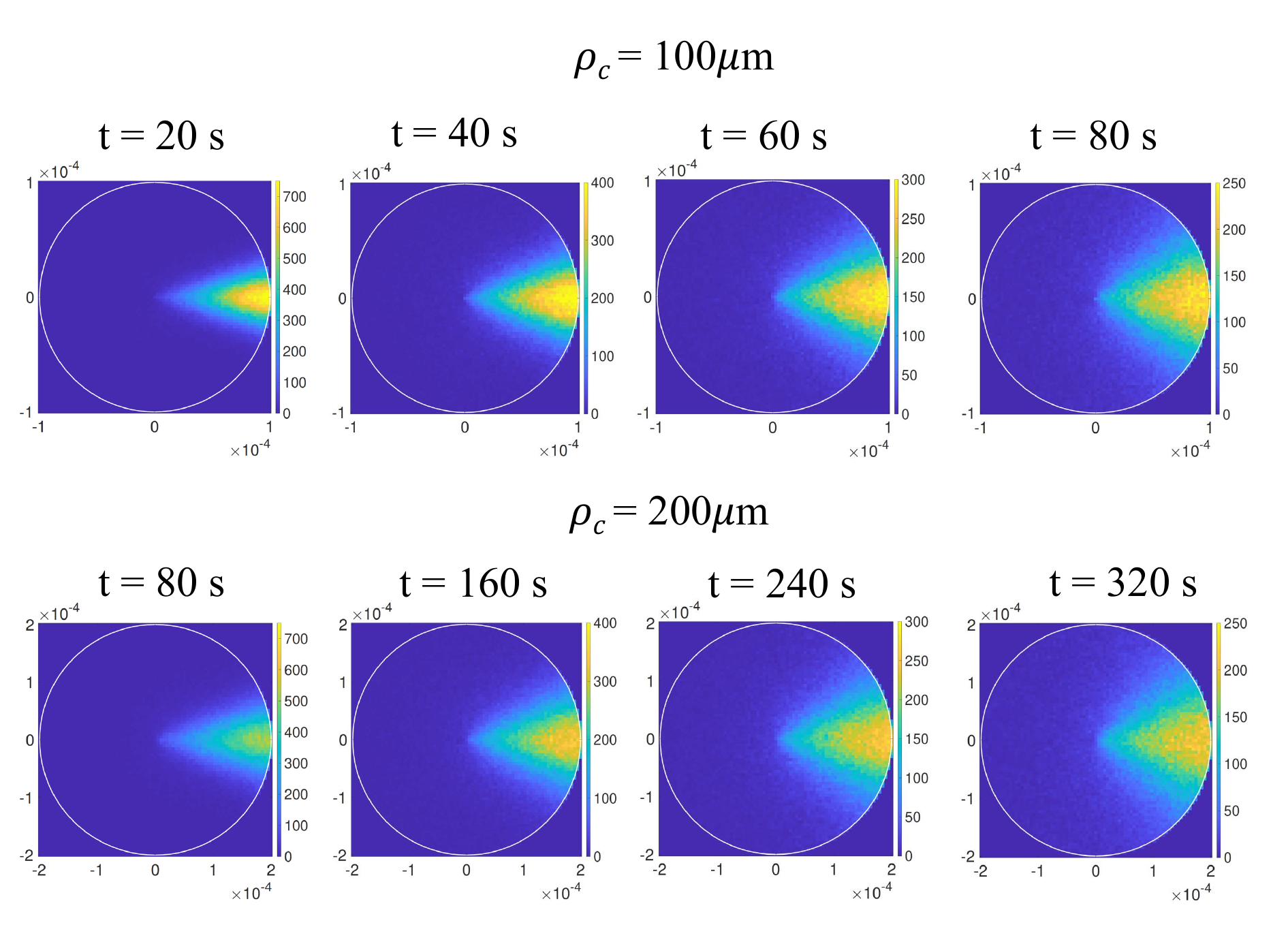}
\caption{2D spatiotemporal profile of anisotropic diffusion organized into two rows corresponding to different biofilm radii and pixel size, $\rho_{c} = \{100, 200\}\,\mu\mathrm{m}$ and pixel dimension $\ell= \{2.5, 5\}\mu$m across, respectively. The top row shows temporal snapshots at $t = \{20, 40, 60, 80\}\,\mathrm{s}$, while the bottom row shows snapshots at $t = \{80, 160, 240, 320\}\,\mathrm{s}$.}
\label{Dimensional Scaling}
\end{figure}

In Fig.~\ref{Dimensional Scaling}, we compare anisotropic diffusion over biofilms of two different sizes, where $D_{\rho}=5\times 10^{-10}\,\text{m}^2 \cdot \text{s}^{-1}$ and $D_{\theta}=5\times 10^{-11}\,\text{m}^2 \cdot \text{s}^{-1}$. Along the top row, we show propagation across a biofilm with radius \(\rho _{c} = 100\,\mu\text{m}\) and observe the local concentrations at times $t = \{20, 40, 60, 80\}\,\mathrm{s}$. Along the bottom row, we show propagation across a biofilm with double the radius \(\rho _{c} = 200\,\mu\text{m}\) and at quadruple the observation times $t = \{80, 160, 240, 320\}\,\mathrm{s}$, but also double the color map pixel element width to $\ell=5\,\mu$m. In both cases, we have the TX placed at $\rho=\rho_c$ and $\theta=0$. Since the results for the two rows appear extremely similar, we can be confident that anisotropic diffusion in our biofilm channel model scales similarly to isotropic diffusion, such that the propagation time increases with the square of the distance. Thus, we consider the smaller geometry with \(\rho _{c} = 100\,\mu\text{m}\) for the remainder of the paper.

\subsection{Comparing the Time-Varying and Static Channels}

We now compare the diffusion dynamics under isotropic, anisotropic, and time-varying conditions. In Fig.~\ref{Spatiotemporal Profile of Iso, Aniso and TVC}, we present color maps that illustrate the spatiotemporal variations induced by isotropic (top row; $D_{\rho}=D_{\theta}=5\times 10^{-10}\,\text{m}^2 \cdot \text{s}^{-1}$), static anisotropic (middle row; $D_{\rho}=5\times 10^{-10}\,\text{m}^2 \cdot \text{s}^{-1}$ and $D_{\theta}=5\times 10^{-11}\,\text{m}^2 \cdot \text{s}^{-1}$), and time-varying anisotropic (bottom row; $D_{\rho}=5\times 10^{-10}\,\text{m}^2 \cdot \text{s}^{-1}$ and $D_{\theta}=D_{\theta}(t)$ as in (\ref{eqn_timevarying})) diffusion, with the transmitter placed at $\rho=\rho_c$ and $\theta=0$. The four columns correspond to snapshots taken at times t = $\{20, 40, 60, 80\}$\,s, respectively, and each column uses the same color map scale. In general, the isotropic diffusion system produces diffusion profiles that are more circularly symmetric about the transmitter. Anisotropic diffusion produces diffusion profiles with much more radial diffusion than axial diffusion, such that there is much less axial spread and much more radial propagation. This is as we expect, since \( D_{\rho} \) is an order of magnitude greater than \( D_{\theta} \). The TVC system exhibits hybrid behavior, transitioning from isotropic to increasingly anisotropic diffusion over time. The profile at $t=20\,$s appears identical to the profile in the isotropic case, as we expect since the initial diffusion parameters are the same. Nevertheless, by $t=80\,$s the profile has a wedge shape that much more closely resembles the fully anisotropic case, except for a less concentrated peak around the transmitter location due to the spreading induced when the AI molecules were released and the system was initially isotropic.

\begin{figure}[t!]
\includegraphics[width=1\linewidth]{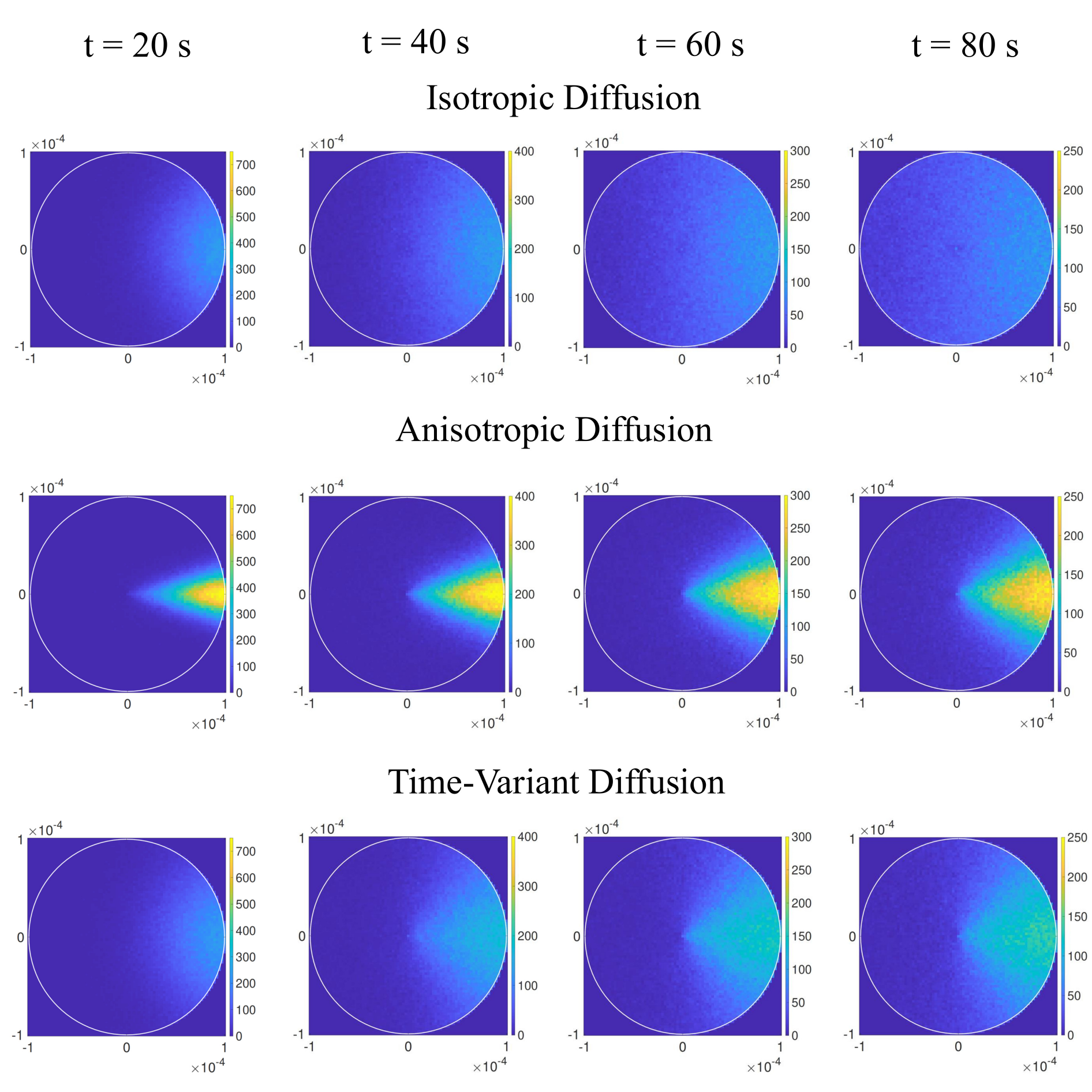}
\caption{2D spatiotemporal profile of isotropic ($D_{\theta}=5\times 10^{-10}\,\text{m}^2 \cdot \text{s}^{-1}$), anisotropic ($D_{\theta}=5\times 10^{-11}\,\text{m}^2 \cdot \text{s}^{-1}$), and time-varying ($D_{\theta}=D_{\theta}(t)$ as in (\ref{eqn_timevarying})) diffusion. In all cases, $D_{\rho}=5\times 10^{-10}\,\text{m}^2 \cdot \text{s}^{-1}$ and the TX is placed at the rightmost edge. Each column corresponds to temporal snapshots at $\mathrm{t}$ = \{20, 40, 60, 80\} s.}
\label{Spatiotemporal Profile of Iso, Aniso and TVC}
\end{figure}

\begin{figure}[t!]
\includegraphics[width=1\linewidth]{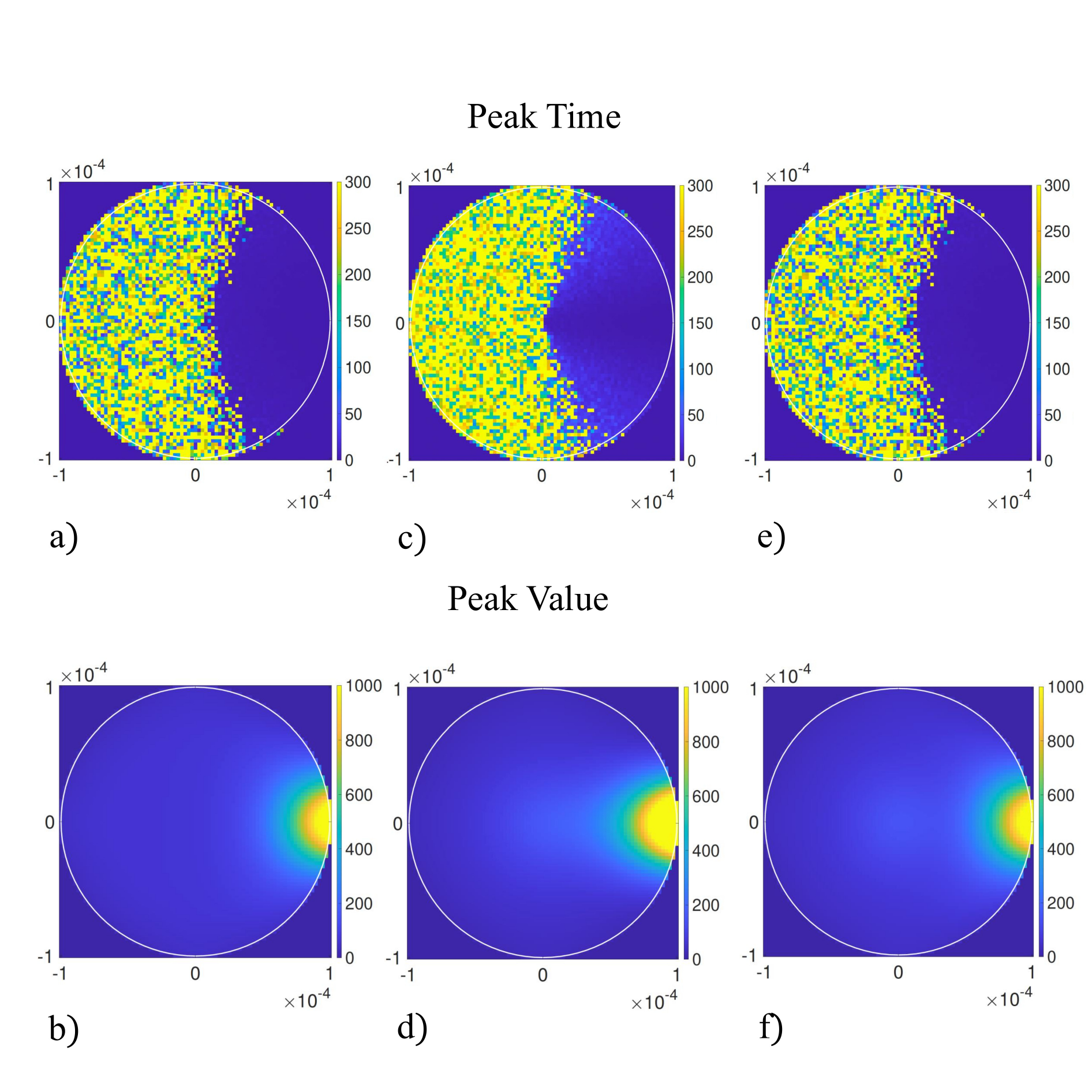}
\caption{2D diffusion profiles under isotropic \{(a), (b)\}, anisotropic \{(c), (d)\}, and time-varying \{(e), (f)\} conditions simulated for 600\,s. The top row shows peak times measured at least 0.3\,s and within 300\,s after molecule release. The bottom row shows the peak values.}
\label{Peak Time and Peak Value}
\end{figure}

In Fig.~\ref{Peak Time and Peak Value}, we present color maps that illustrate the peak times and peak values under isotropic \{(a), (b)\}, anisotropic \{(c), (d)\}, and time-varying \{(e), (f)\} diffusion conditions, again with the TX on the boundary at $\rho=\rho_c$ and $\theta=0$. The top row measures the time (in seconds) at which the peak number of AIs was observed within each pixel, after an initial wait of 0.3\,s to give AIs a chance to initially spread away from the TX. The bottom row measures the peak number of AI molecules observed within each pixel. In all three cases, the system was simulated for $600\,$s but we only indicate in the top row where the peak was observed within the first $300\,$s, to improve the visibility and resolution of the peak time in the region closer to the transmitter. The fully anisotropic system exhibits a larger region with higher peak values when compared with both the isotropic and time-varying anisotropic cases. In particular, higher peaks are observed along the radial direction towards the center of the biofilm, demonstrating how propagation in the anisotropic system is more penetrating even though the radial diffusion coefficient $D_{\rho}$ is the same. In the time-varying system, the peak behavior near the TX is very similar to that in the isotropic system, which reflects the fact that the time-varying system was initially isotropic and the peak signal near the TX will be soon after transmission. Nevertheless, close inspection of the peak values near the center of the biofilm reveals that they are measurably larger for the time-varying system than for the isotropic system, which is consistent with the time-varying system becoming gradually more anisotropic.

\begin{table}[t!]
\centering
\caption{Comparison of Predicted and Observed Peak Values for Anisotropic Diffusion, Measured at Three Arbitrary Locations}
\label{tab:peak_values}
\resizebox{\columnwidth}{!}{%
\begin{tabular}{@{}ccc@{}}
\toprule
\textbf{Predicted Peak Value} & \textbf{Average Observed Peak Value} & \textbf{Percentage Deviation (\%)} \\
\midrule
610 & 590 & 3.39 \\
730 & 710 & 2.82 \\
560 & 530 & 5.66 \\

\bottomrule
\end{tabular}%
\label{Table II}
}
\end{table}

In Table~\ref{Table II}, we compare a small selection of the predicted peak values with those observed in the static anisotropic channel simulations at the corresponding peak times evaluated from (\ref{equation peak}). For this analysis, three arbitrary locations in the anisotropic diffusion system were selected, with results at each location averaged over ten runs. The locations were chosen from within the green-shaded region in Fig.~\ref{Peak Time and Peak Value}(d), representing observed peak molecule counts between 500 and 750 molecules. The predicted peak value at each location was obtained by setting the derivative in (\ref{equation peak}) to zero, solving for the peak time, substituting this time into the channel impulse response in (\ref{equation 4}), and scaling for both the number of molecules released and the pixel size. At each location, we measure the deviation in the predicted peak value from the observed peak value, which was less than 6\,\% at all three locations considered.

\subsection{Parameter Sensitivity of Mutual Information }

\begin{figure}[t!]
\includegraphics[width=1\linewidth]{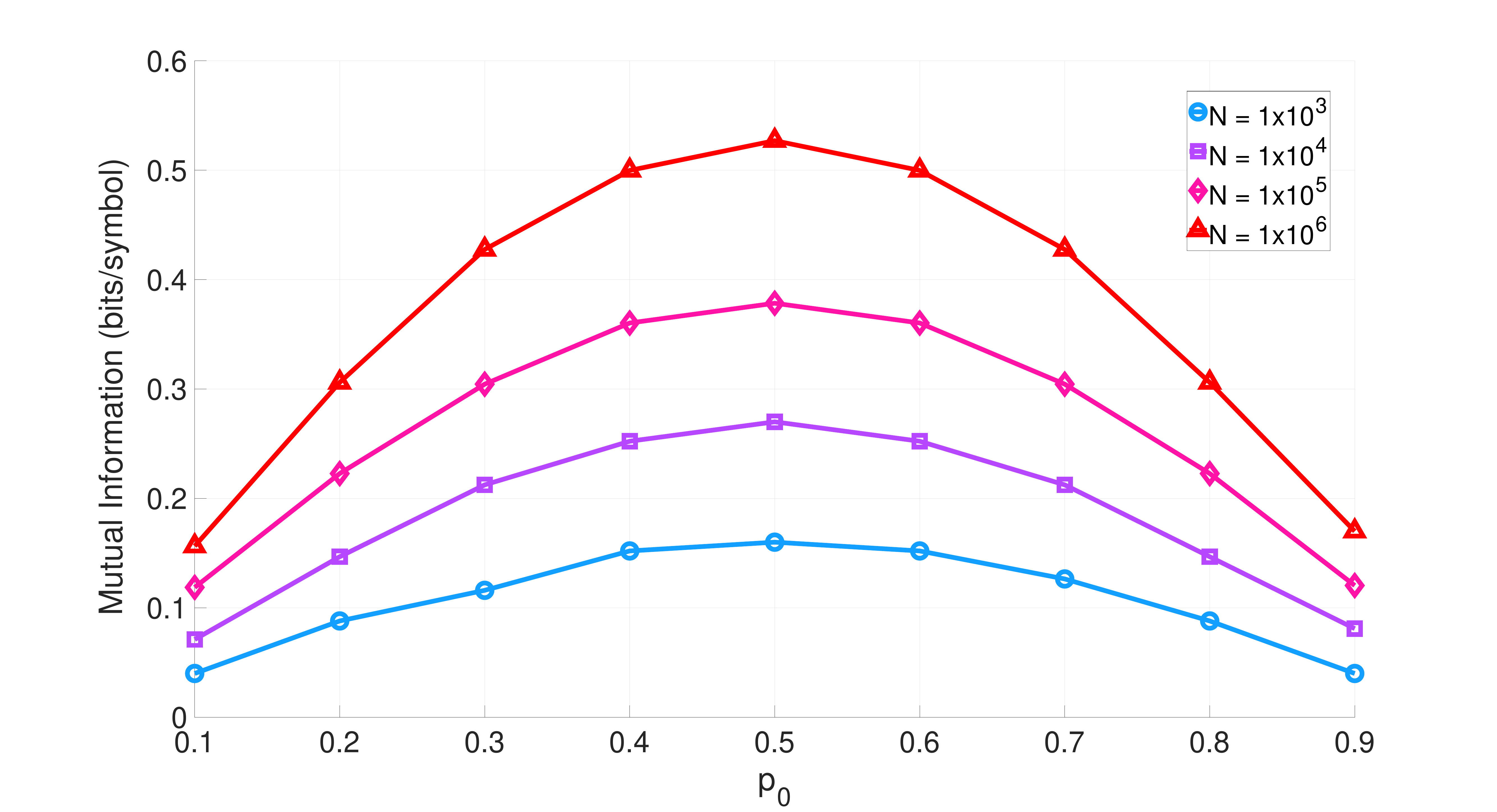}

\caption{The relationship of the MI and  $p_0$ for different number of AIs ($N$).}
\label{MI_N}
\end{figure}

\begin{figure}[t!]
\includegraphics[width=1\linewidth]{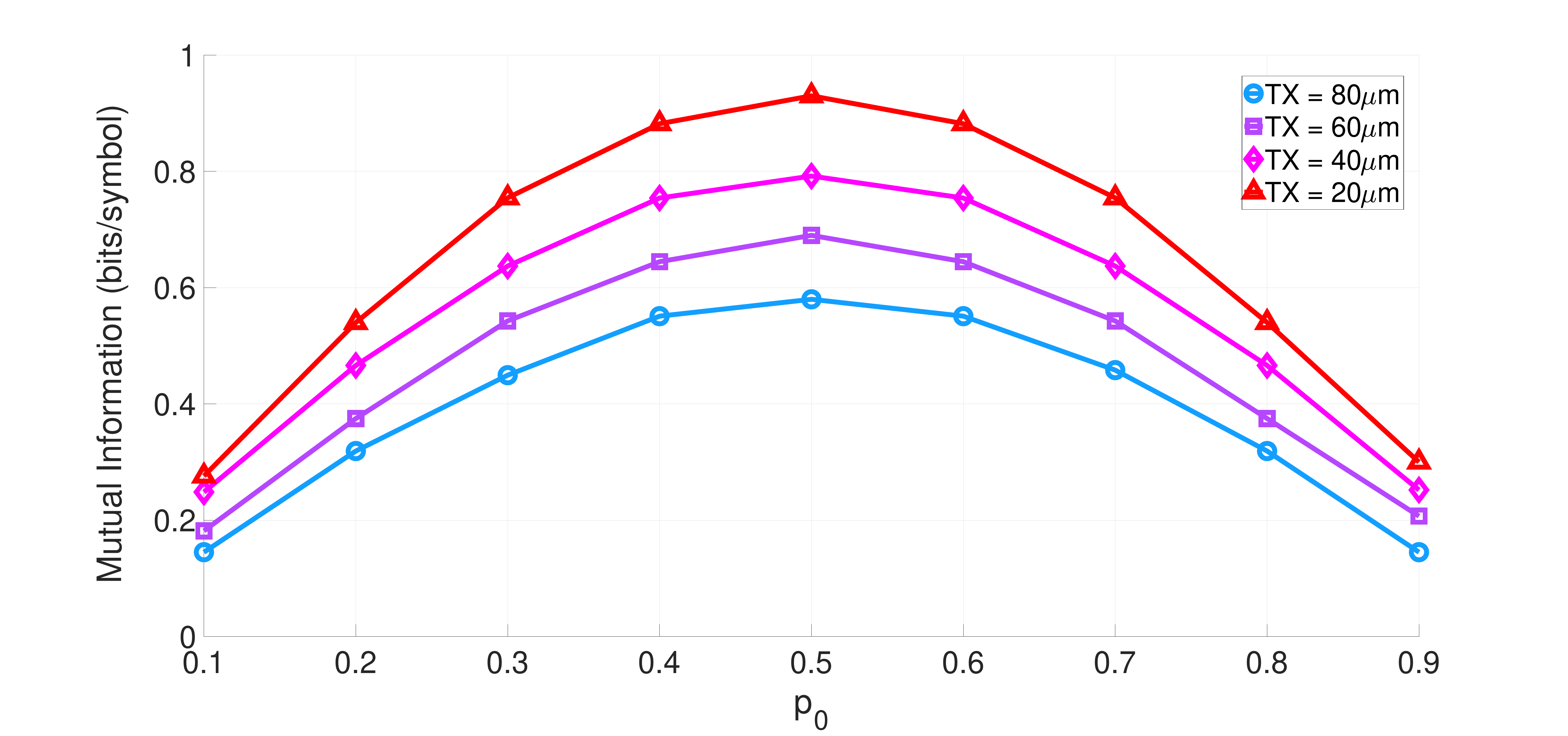}

\caption{Relationship between MI and $p_0$ for different TX--RX distances. TX locations were set at $\rho = \{20, 40, 60, 80\}\,\mu\text{m}$ with $\theta = 0$ rad, and observations were made at the RX positioned at (0 $\mu$m, 0 rad).}
\label{MI_TX}
\end{figure}

\begin{figure}[t!]
\includegraphics[width=1\linewidth]{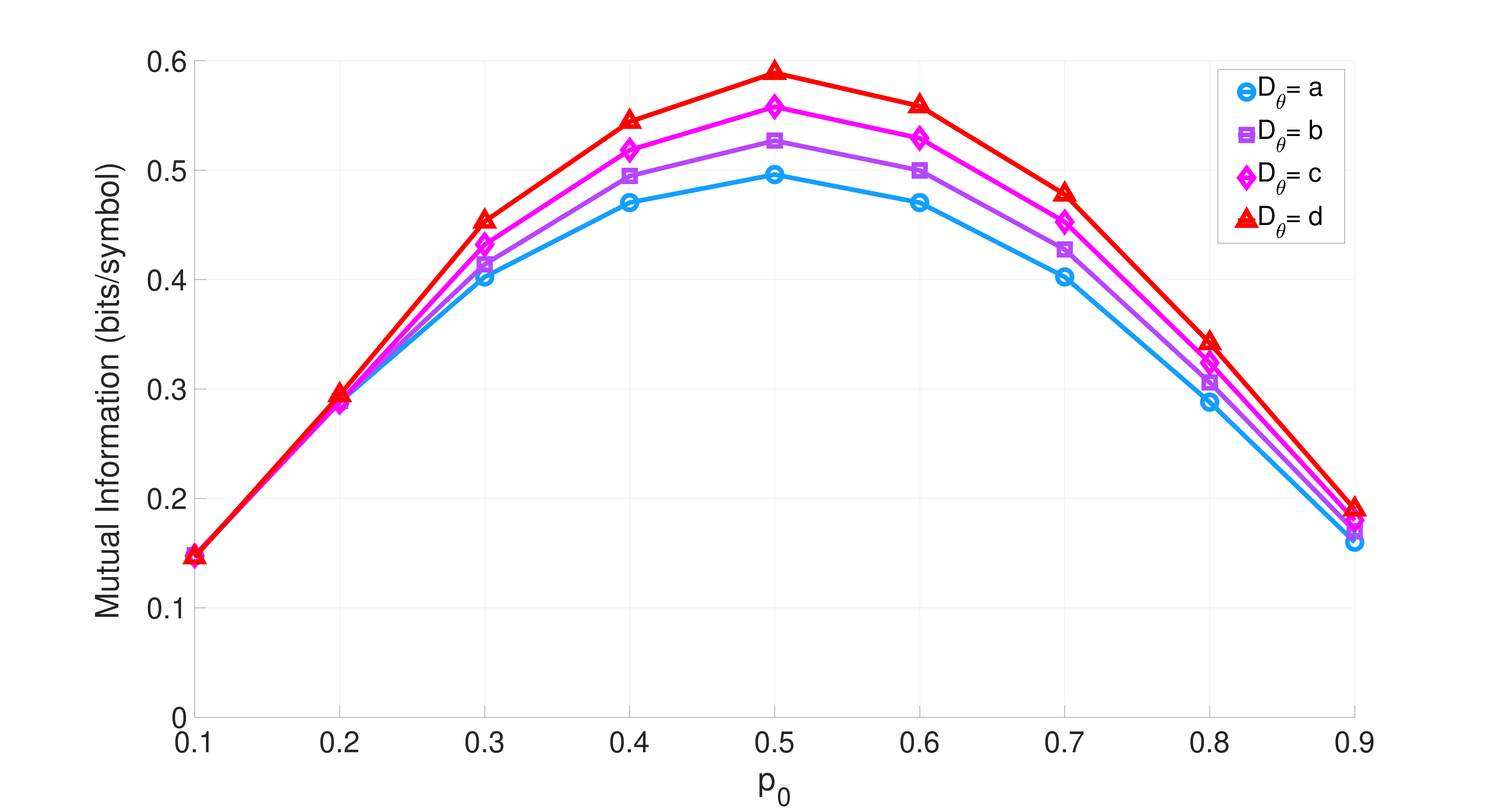}

\caption{The relationship between MI and $p_0$ for different 
$D_{\theta}$ values in time-varying scenarios of anisotropy. Each curve corresponds to a sequence of four time intervals at $t=\{20,40,60,80\}\,\text{s}$, where $D_{\theta}$ is updated at each interval.
The sets of $D_{\theta}$ values used are as follows: 
(a) $\{5 \times 10^{-10}, 4 \times 10^{-10}, 3 \times 10^{-10}, 2 \times 10^{-10}\} \, \text{m}^2 \cdot \text{s}^{-1}$,  
(b) $\{5 \times 10^{-10}, 5 \times 10^{-11}, 3 \times 10^{-11}, 1 \times 10^{-11}\} \, \text{m}^2 \cdot \text{s}^{-1}$,  
(c) $\{5 \times 10^{-10}, 7 \times 10^{-11}, 5 \times 10^{-11}, 3 \times 10^{-11}\} \, \text{m}^2 \cdot \text{s}^{-1}$,  
(d) $\{5 \times 10^{-10}, 5 \times 10^{-11}, 2 \times 10^{-11}, 1 \times 10^{-11}\} \, \text{m}^2 \cdot \text{s}^{-1}$.
}
\label{MI_D}

\end{figure}

\begin{figure}[t!]
\includegraphics[width=1\linewidth]{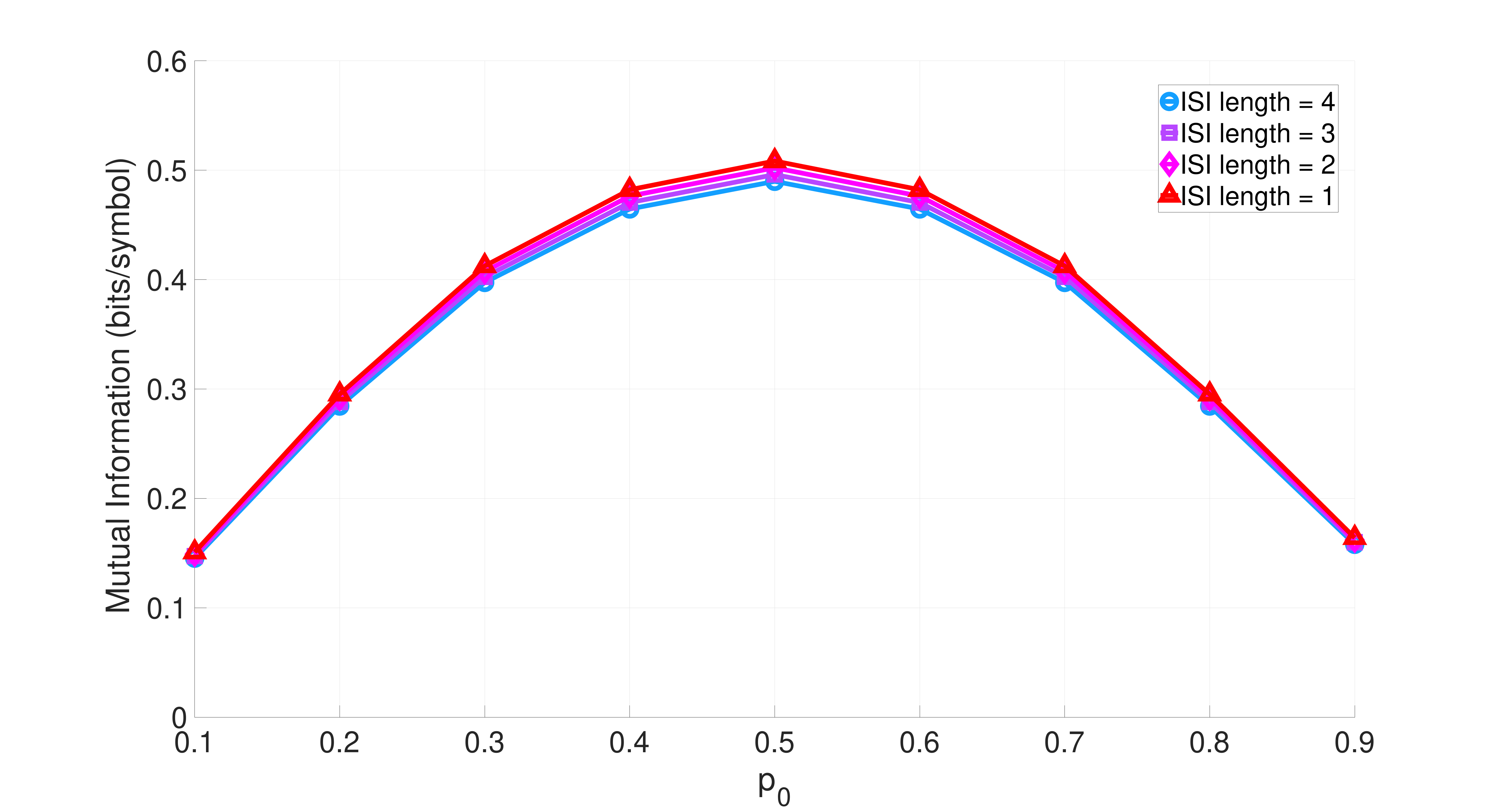}

\caption{The relationship of MI and  $p_0$ for different
ISI lenghts.}
\label{MI_ISI}
\end{figure}

To measure the mutual information in the TVC system, we place the RX at the center and transmit a sequence of $L=5$ symbols with a symbol interval of $T_s=20\,$s. In particular, we vary the number of AI molecules released, the precise TX-RX placement (by varying the location of the TX), the diffusion parameters, and the specific transmission interval considered. The corresponding results, averaged over 500 trials, are presented in Figs.~\ref{MI_N},~\ref{MI_TX},~\ref{MI_D}, and~\ref{MI_ISI}, respectively. In all simulations, we do not include the observation of the first symbol (i.e., $Y_0$); this symbol skews the computation since there is no ISI. Instead, unless otherwise noted, we average the MI calculated for all later symbols $\{Y_1,\ldots,Y_5\}$.

In Fig.~\ref{MI_N}, we illustrate the impact of the symbol probability $p_0$ and the number of AI molecules $N$ released for symbol ``1'' on the MI. A logarithmic relationship with respect to the number of AIs is observed; increasing $N$ results in greater mutual information. The highest MI peak is observed at $p_0 = 0.5$ with $N = 10^{6}$ molecules, whereas the smallest MI peak is observed with $N = 10^{3}$ molecules, where the MI is approximately $0.18$ bits per symbol.

In Fig.~\ref{MI_TX}, the impact of TX--RX placement on MI is examined, while Fig.~\ref{MI_D} investigates the effect of varying the strength of anisotropy within the TVC. Both figures showcase an initial increase followed by a decrease in MI with an increase in symbol probability. When \( p_0 = 0.5 \), the MI reaches its maximum value. As expected, in Fig.~\ref{MI_TX}, with a decrease in TX-RX placement distance, there is an increase in MI as signals reach the RX more reliably. In Fig.~\ref{MI_D}, the MI is generally higher with a greater degree of anisotropy. In effect, the diffusion transforms from 2D to 1D diffusion as the system becomes more anisotropic and transmission becomes more reliable. Another observation is that all plots in Fig.~\ref{MI_D} are asymmetric with respect to symbol probability \( p_0 \), and this asymmetry becomes more pronounced with a greater degree of anisotropy.

In Fig.~\ref{MI_ISI}, we illustrate the effect of symbol probability \( p_0 \)  and ISI on MI for each individual symbol $Y_s$ (i.e., we do not average the MI calculation over the last 4 symbols). The effects of ISI can persist since the AIs released in one symbol interval can be observed in a later interval. The ISI length refers to the number of previous symbols, so symbol $Y_s$ has an ISI length of $s$ and the maximum is $s=4$. While the differences in Fig.~\ref{MI_ISI} are subtle, there is an observable decrease in the MI as ISI accumulates in the system, which is sufficient to offset the gains that we would expect from the channel becoming more anisotropic over the timescale of the transmitted symbols.

\subsection{Discussion}

Collectively, our results demonstrate the changes in a biofilm's signal propagation characteristics as it matures and gradually becomes more anisotropic. Even with a simple time-varying model, we see in Fig.~\ref{Spatiotemporal Profile of Iso, Aniso and TVC} how spatiotemporal profiles become more anisotropic, even if the system was originally isotropic when AIs were transmitted. Fig.~\ref{MI_D} suggests that an increase in anisotropy also leads to an increase in information that can be transmitted from the edge of the biofilm to the center. It is interesting that directionality is enhanced even when the anisotropy develops gradually during transmission. This suggests that biofilm maturation could be guided to favor anisotropy in order to support the coordination of activities between the edge of a biofilm, where there is greater access to external molecules but higher exposure to dangers, and the center of the biofilm, which has better protection from external forces but is more resource-scarce.

Similar to what we proposed in \cite{paramalingam2025anisotropic}, experimental single-particle tracking techniques would be helpful to gain further insight into AI diffusion dynamics during biofilm maturation and to validate or update our proposed model. Such experimental data would be invaluable for a more precise characterization of the effective diffusion coefficient in a maturing biofilm and the extent to which biofilms develop and exhibit anisotropy during this stage.

\section{Conclusion}

This paper extended a previously-established 2D anisotropic diffusion model for molecular communication in a biofilm. The extension enabled the anisotropic diffusion parameters to vary over time in order to approximate the changes in the signal propagation dynamics as the biofilm matures. The time-varying channel (TVC) was characterized by a coherence time over which the diffusion parameters were assumed to be fixed. Our simulation results demonstrated how the spatiotemporal profile of the TVC acted as a hybrid of the isotropic and anisotropic systems, reflecting how the TVC became gradually more anisotropic. We also presented how diffusion in an anisotropic systems scales similarly to isotropic systems, where the propagation time increases with the square of the propagation distance. We characterized the peak observed concentration and how it is impacted by the TVC. Finally, we measured the mutual information (MI) within communication links of a TVC system and how the MI increases with the number of AIs transmitted, an increase in anisotropy, a decrease in transmission distance, and a decrease in ISI length.

Overall, our results reflect the impact of changing anisotropy in biofilm communication links. Future works can consider a more detailed time-varying model, communication between more than two nodes (e.g., via relaying or broadcast signaling), and more precisely characterizing how bacterial messages can propagate through a biofilm.

\section*{Acknowledgment}

The authors would like to thank Dr. Freya Harrison and Dr. Tara Schiller for their constructive feedback and support, particularly on the biological details of this work.

\bibliographystyle{IEEEtran}
\bibliography{Reference}

\begin{thebibliography}{10}
\providecommand{\url}[1]{#1}
\csname url@samestyle\endcsname
\providecommand{\newblock}{\relax}
\providecommand{\bibinfo}[2]{#2}
\providecommand{\BIBentrySTDinterwordspacing}{\spaceskip=0pt\relax}
\providecommand{\BIBentryALTinterwordstretchfactor}{4}
\providecommand{\BIBentryALTinterwordspacing}{\spaceskip=\fontdimen2\font plus
\BIBentryALTinterwordstretchfactor\fontdimen3\font minus
  \fontdimen4\font\relax}
\providecommand{\BIBforeignlanguage}[2]{{%
\expandafter\ifx\csname l@#1\endcsname\relax
\typeout{** WARNING: IEEEtran.bst: No hyphenation pattern has been}%
\typeout{** loaded for the language `#1'. Using the pattern for}%
\typeout{** the default language instead.}%
\else
\language=\csname l@#1\endcsname
\fi
#2}}
\providecommand{\BIBdecl}{\relax}
\BIBdecl

\bibitem{hall2004bacterial}
L.~Hall-Stoodley, J.~W. Costerton, and P.~Stoodley, ``Bacterial biofilms: from
  the natural environment to infectious diseases,'' \emph{Nature Reviews
  Microbiology}, vol.~2, no.~2, pp. 95--108, 2004.

\bibitem{costerton1999bacterial}
J.~W. Costerton, P.~S. Stewart, and E.~P. Greenberg, ``Bacterial biofilms: a
  common cause of persistent infections,'' \emph{Science}, vol. 284, no. 5418,
  pp. 1318--1322, 1999.

\bibitem{vestby2020bacterial}
L.~K. Vestby, T.~Gr{\o}nseth, R.~Simm, and L.~L. Nesse, ``Bacterial biofilm and
  its role in the pathogenesis of disease,'' \emph{Antibiotics}, vol.~9, no.~2,
  p.~59, 2020.

\bibitem{mukherjee2019bacterial}
S.~Mukherjee and B.~L. Bassler, ``Bacterial quorum sensing in complex and
  dynamically changing environments,'' \emph{Nature Reviews Microbiology},
  vol.~17, no.~6, pp. 371--382, 2019.

\bibitem{quan2022water}
K.~Quan, J.~Hou, Z.~Zhang, Y.~Ren, B.~W. Peterson, H.-C. Flemming, C.~Mayer,
  H.~J. Busscher, and H.~C. van~der Mei, ``Water in bacterial biofilms: pores
  and channels, storage and transport functions,'' \emph{Critical Reviews in
  Microbiology}, vol.~48, no.~3, pp. 283--302, 2022.

\bibitem{wilking2013liquid}
J.~N. Wilking, V.~Zaburdaev, M.~De~Volder, R.~Losick, M.~P. Brenner, and D.~A.
  Weitz, ``Liquid transport facilitated by channels in \textit{Bacillus
  subtilis} biofilms,'' \emph{Proceedings of the National Academy of Sciences},
  vol. 110, no.~3, pp. 848--852, 2013.

\bibitem{rooney2020intra}
L.~M. Rooney, W.~B. Amos, P.~A. Hoskisson, and G.~McConnell, ``Intra-colony
  channels in \textit{E. coli} function as a nutrient uptake system,''
  \emph{The ISME journal}, vol.~14, no.~10, pp. 2461--2473, 2020.

\bibitem{rosero2021microbial}
G.~Rosero-Chasoy, R.~M. Rodr{\'\i}guez-Jasso, C.~N. Aguilar, G.~Buitr{\'o}n,
  I.~Chairez, and H.~A. Ruiz, ``Microbial co-culturing strategies for the
  production high value compounds, a reliable framework towards sustainable
  biorefinery implementation--an overview,'' \emph{Bioresource Technology},
  vol. 321, p. 124458, 2021.

\bibitem{bjarnsholt2007quorum}
T.~Bjarnsholt and M.~Givskov, ``Quorum-sensing blockade as a strategy for
  enhancing host defences against bacterial pathogens,'' \emph{Philosophical
  Transactions of the Royal Society B: Biological Sciences}, vol. 362, no.
  1483, pp. 1213--1222, 2007.

\bibitem{paluch2020prevention}
E.~Paluch, J.~Rewak-Soroczy{\'n}ska, I.~Jedrusik, E.~Mazurkiewicz, and
  K.~Jermakow, ``Prevention of biofilm formation by quorum quenching,''
  \emph{Applied Microbiology and Biotechnology}, vol. 104, pp. 1871--1881,
  2020.

\bibitem{kannan2018mathematical}
R.~E. Kannan and S.~Saini, ``Mathematical modelling of quorum sensing in
  bacteria,'' \emph{INAE Letters}, vol.~3, pp. 175--187, 2018.

\bibitem{perez2016mathematical}
J.~P{\'e}rez-Vel{\'a}zquez, M.~G{\"o}lgeli, and R.~Garc{\'\i}a-Contreras,
  ``Mathematical modelling of bacterial quorum sensing: a review,''
  \emph{Bulletin of Mathematical Biology}, vol.~78, pp. 1585--1639, 2016.

\bibitem{li2020quantitative}
X.~Li, H.~Qi, X.-C. Zhang, F.~Xu, Z.-Y. Yin, S.-Y. Huang, Z.-S. Wang, and J.-W.
  Shuai, ``Quantitative modeling of bacterial quorum sensing dynamics in time
  and space,'' \emph{Chinese Physics B}, vol.~29, no.~10, p. 108702, 2020.

\bibitem{mattei2018continuum}
M.~Mattei, L.~Frunzo, B.~D’acunto, Y.~Pechaud, F.~Pirozzi, and G.~Esposito,
  ``Continuum and discrete approach in modeling biofilm development and
  structure: a review,'' \emph{Journal of Mathematical Biology}, vol.~76, pp.
  945--1003, 2018.

\bibitem{paramalingam2025anisotropic}
Y.~Paramalingam, H.~Arjmandi, F.~Harrison, T.~Schiller, and A.~Noel,
  ``Anisotropic diffusion model of communication in 2{D} biofilm,'' \emph{IEEE
  Transactions on Molecular, Biological, and Multi-Scale Communications}, 2025.

\bibitem{farsad2016comprehensive}
N.~Farsad, H.~B. Yilmaz, A.~Eckford, C.-B. Chae, and W.~Guo, ``A comprehensive
  survey of recent advancements in molecular communication,'' \emph{IEEE
  Communications Surveys \& Tutorials}, vol.~18, no.~3, pp. 1887--1919, 2016.

\bibitem{van2012anisotropic}
A.~Van~Wey, A.~Cookson, T.~Soboleva, N.~Roy, W.~McNabb, A.~Bridier,
  R.~Briandet, and P.~Shorten, ``Anisotropic nutrient transport in
  three-dimensional single species bacterial biofilms,'' \emph{Biotechnology
  and Bioengineering}, vol. 109, no.~5, pp. 1280--1292, 2012.

\bibitem{noel2017effect}
A.~Noel, Y.~Fang, N.~Yang, D.~Makrakis, and A.~W. Eckford, ``Effect of local
  population uncertainty on cooperation in bacteria,'' in \emph{2017 IEEE
  Information Theory Workshop (ITW)}.\hskip 1em plus 0.5em minus 0.4em\relax
  IEEE, 2017, pp. 334--338.

\bibitem{martins2022microfluidic}
D.~P. Martins, M.~T. Barros, B.~J. O’Sullivan, I.~Seymour, A.~O’Riordan,
  L.~Coffey, J.~B. Sweeney, and S.~Balasubramaniam, ``Microfluidic-based
  bacterial molecular computing on a chip,'' \emph{IEEE Sensors Journal},
  vol.~22, no.~17, pp. 16\,772--16\,784, 2022.

\bibitem{einolghozati2013relaying}
A.~Einolghozati, M.~Sardari, and F.~Fekri, ``Relaying in diffusion-based
  molecular communication,'' in \emph{2013 IEEE International Symposium on
  Information Theory}.\hskip 1em plus 0.5em minus 0.4em\relax IEEE, 2013, pp.
  1844--1848.

\bibitem{balasubramaniam2023realizing}
S.~Balasubramaniam, S.~Somathilaka, S.~Sun, A.~Ratwatte, and M.~Pierobon,
  ``Realizing molecular machine learning through communications for biological
  {AI},'' \emph{IEEE Nanotechnology Magazine}, 2023.

\bibitem{cobo2010bacteria}
L.~C. Cobo and I.~F. Akyildiz, ``Bacteria-based communication in
  nanonetworks,'' \emph{Nano Communication Networks}, vol.~1, no.~4, pp.
  244--256, 2010.

\bibitem{unluturk2015genetically}
B.~D. Unluturk, A.~O. Bicen, and I.~F. Akyildiz, ``Genetically engineered
  bacteria-based biotransceivers for molecular communication,'' \emph{IEEE
  Transactions on Communications}, vol.~63, no.~4, pp. 1271--1281, 2015.

\bibitem{tissera2020bio}
P.~S.~S. Tissera and S.~Choe, ``Bio-inspired quorum sensing-based nanonetwork
  synchronization using birth-death growth model,'' \emph{IEEE Transactions on
  Communications}, vol.~68, no.~10, pp. 6263--6275, 2020.

\bibitem{einolghozati2013design}
A.~Einolghozati, M.~Sardari, and F.~Fekri, ``Design and analysis of wireless
  communication systems using diffusion-based molecular communication among
  bacteria,'' \emph{IEEE Transactions on Wireless Communications}, vol.~12,
  no.~12, pp. 6096--6105, 2013.

\bibitem{fang2020characterization}
Y.~Fang, A.~Noel, A.~W. Eckford, N.~Yang, and J.~Guo, ``Characterization of
  cooperators in quorum sensing with 2d molecular signal analysis,'' \emph{IEEE
  Transactions on Communications}, vol.~69, no.~2, pp. 799--816, 2020.

\bibitem{abadal2012quorum}
S.~Abadal, I.~Llatser, E.~Alarc{\'o}n, and A.~Cabellos-Aparicio, ``Quorum
  sensing-enabled amplification for molecular nanonetworks,'' in \emph{2012
  IEEE International Conference on Communications (ICC)}.\hskip 1em plus 0.5em
  minus 0.4em\relax IEEE, 2012, pp. 6162--6166.

\bibitem{tissera2019quorum}
P.~S. Tissera, S.~Choe, and R.~Punmiya, ``Quorum sensing-based nanonetwork
  synchronization,'' \emph{IEEE Wireless Communications Letters}, vol.~8,
  no.~3, pp. 893--896, 2019.

\bibitem{li2015evaluation}
F.~Li, L.~Lin, C.~Yang, and M.~Ma, ``Evaluation of molecular oscillation for
  nanonetworks based on quorum sensing,'' in \emph{2015 1st Workshop on
  Nanotechnology in Instrumentation and Measurement (NANOFIM)}.\hskip 1em plus
  0.5em minus 0.4em\relax IEEE, 2015, pp. 233--237.

\bibitem{martins2018molecular}
D.~P. Martins, K.~Leetanasaksakul, M.~T. Barros, A.~Thamchaipenet, W.~Donnelly,
  and S.~Balasubramaniam, ``Molecular communications pulse-based jamming model
  for bacterial biofilm suppression,'' \emph{IEEE Transactions on
  Nanobioscience}, vol.~17, no.~4, pp. 533--542, 2018.

\bibitem{michelusi2016queuing}
N.~Michelusi, J.~Boedicker, M.~Y. El-Naggar, and U.~Mitra, ``Queuing models for
  abstracting interactions in bacterial communities,'' \emph{IEEE Journal on
  Selected Areas in Communications}, vol.~34, no.~3, pp. 584--599, 2016.

\bibitem{martins2016using}
D.~P. Martins, M.~T. Barros, and S.~Balasubramaniam, ``Using competing
  bacterial communication to disassemble biofilms,'' in \emph{Proceedings of
  the 3rd ACM International Conference on Nanoscale Computing and
  Communication}, 2016, pp. 1--6.

\bibitem{gulec2023stochastic}
F.~Gulec and A.~W. Eckford, ``A stochastic biofilm disruption model based on
  quorum sensing mimickers,'' \emph{IEEE Transactions on Molecular, Biological
  and Multi-Scale Communications}, 2023.

\bibitem{flemming2010biofilm}
H.-C. Flemming and J.~Wingender, ``The biofilm matrix,'' \emph{Nature reviews
  microbiology}, vol.~8, no.~9, pp. 623--633, 2010.

\bibitem{kuchma2000surface}
S.~L. Kuchma and G.~A. O'Toole, ``Surface-induced and biofilm-induced changes
  in gene expression,'' \emph{Current opinion in biotechnology}, vol.~11,
  no.~5, pp. 429--433, 2000.

\bibitem{ahmadzadeh2018stochastic}
A.~Ahmadzadeh, V.~Jamali, and R.~Schober, ``Stochastic channel modeling for
  diffusive mobile molecular communication systems,'' \emph{IEEE Transactions
  on Communications}, vol.~66, no.~12, pp. 6205--6220, 2018.

\bibitem{8254237}
------, ``Statistical analysis of time-variant channels in diffusive mobile
  molecular communications,'' in \emph{GLOBECOM 2017 - 2017 IEEE Global
  Communications Conference}, 2017, pp. 1--7.

\bibitem{cao2019diffusive}
T.~N. Cao, A.~Ahmadzadeh, V.~Jamali, W.~Wicke, P.~L. Yeoh, J.~Evans, and
  R.~Schober, ``Diffusive mobile mc for controlled-release drug delivery with
  absorbing receiver,'' in \emph{ICC 2019-2019 IEEE International Conference on
  Communications (ICC)}.\hskip 1em plus 0.5em minus 0.4em\relax IEEE, 2019, pp.
  1--7.

\bibitem{rouzegar2019diffusive}
S.~R. Rouzegar and U.~Spagnolini, ``Diffusive mimo molecular communications:
  Channel estimation, equalization, and detection,'' \emph{IEEE Transactions on
  Communications}, vol.~67, no.~7, pp. 4872--4884, 2019.

\bibitem{feng2018resource}
L.~Feng, K.~S. Kwak, and Q.~Yang, ``Resource allocation for time-variant
  channels in the nano-communication networks,'' in \emph{2018 International
  Conference on Information and Communication Technology Convergence
  (ICTC)}.\hskip 1em plus 0.5em minus 0.4em\relax IEEE, 2018, pp. 667--671.

\bibitem{trinh2022molecular}
D.~P. Trinh, Y.~Jeong, C.-B. Chae, and S.-H. Kim, ``Molecular communication in
  inhomogeneous diffusion channels,'' \emph{IEEE Wireless Communications
  Letters}, vol.~11, no.~9, pp. 1975--1979, 2022.

\bibitem{lin2012signal}
W.-A. Lin, Y.-C. Lee, P.-C. Yeh, and C.-h. Lee, ``Signal detection and isi
  cancellation for quantity-based amplitude modulation in diffusion-based
  molecular communications,'' in \emph{2012 IEEE Global Communications
  Conference (GLOBECOM)}.\hskip 1em plus 0.5em minus 0.4em\relax IEEE, 2012,
  pp. 4362--4367.

\bibitem{alberghini2009consequences}
S.~Alberghini, E.~Polone, V.~Corich, M.~Carlot, F.~Seno, A.~Trovato, and
  A.~Squartini, ``Consequences of relative cellular positioning on quorum
  sensing and bacterial cell-to-cell communication,'' \emph{FEMS Microbiology
  Letters}, vol. 292, no.~2, pp. 149--161, 2009.

\bibitem{Noel2013b}
A.~Noel, K.~C. Cheung, and R.~Schober, ``Using dimensional analysis to assess
  scalability and accuracy in molecular communication,'' in \emph{Proc. IEEE
  ICC MoNaCom}, Jun. 2013, pp. 818--823.

\end{thebibliography}

\end{document}